\documentclass[twocolumn,preprintnumbers,amsmath,amssymb]{revtex4}
\usepackage{dcolumn}
\usepackage{bm}
\usepackage{graphicx,subfigure,xcolor}
\usepackage{braket}
\usepackage[normalem]{ulem}

\usepackage{comment}
\usepackage{xcolor}
\usepackage{amsmath}
\usepackage{amssymb}
\usepackage{eucal}
\usepackage{mathrsfs}
\usepackage{amsthm}
\usepackage{epstopdf}
\usepackage{textcomp}
\usepackage{braket}
\usepackage{pdfpages}
\usepackage[utf8]{inputenc}

\newcommand{\wo}{\omega_0}
\newcommand{\wk}{\omega_k}
\newcommand{\wkp}{\omega_{k'}}

\newcommand{\ewo}{e^{i\omega_0 t}}
\newcommand{\ewom}{e^{-i\omega_0 t}}

\newcommand{\ewkm}{e^{-i\omega_k t}}

\newcommand{\ekr}{e^{i{\bf k}\cdot{\bf r}}}
\newcommand{\ekrm}{e^{-i{\bf k}\cdot{\bf r}}}
\newcommand{\ekri}{e^{i{\bf k}\cdot{\bf r}_i}}
\newcommand{\ekrmi}{e^{-i{\bf k}\cdot{\bf r}_i}}

\newcommand{\sAB}{\sum_{i=A,B}}

\newcommand{\sqAB}{\sum_{q=A,B}}
\newcommand{\ssAB}{\sum_{s=A,B}}
\newcommand{\skj}{\sum_{{\bf k}j}}
\newcommand{\skjp}{\sum_{{\bf k'}j'}}

\newcommand{\akjd}{a^{\dag}_{{\bf k}j}}
\newcommand{\akj}{a_{{\bf k}j}}
\newcommand{\akjdp}{a^{\dag}_{{\bf k'}j'}}
\newcommand{\akjp}{a_{{\bf k'}j'}}
\newcommand{\ekj}{{\bf e}_{{\bf k}j}}
\newcommand{\ekjp}{{\bf e}_{{\bf k'}j'}}

\def\bk{{\bf k}}
\def\bkp{{\bf k}'}

\def\br{{\mathbf r}}
\def\bR{{\mathbf R}}

\newcommand{\bmu}{\boldsymbol\mu}

\newcommand{\vac}{\vert vac\rangle}

\begin{document}

\title{Time-dependent resonance interaction energy between two entangled atoms under non-equilibrium conditions}

\author{Giuseppe Fiscelli $^{1,2}$}

\author{Roberta Palacino$^{3}$}

\author{Roberto Passante$^{1,2}$\footnote{roberto.passante@unipa.it}}

\author{Lucia Rizzuto$^{1,2}$}

\author{Salvatore Spagnolo$^{1,4}$}

\author{Wenting Zhou$^{5}$}

\affiliation{$^1$ Universit\`{a} degli Studi di Palermo, Dipartimento di Fisica e Chimica, Via Archirafi 36, I-90123 Palermo, Italia}
\affiliation{$2$ INFN, Laboratori Nazionali del Sud, I-95123 Catania, Italy}
\affiliation{$^3$ SUPA, School of Physics and Astronomy, University of St. Andrews, St. Andrews KY16 9SS, United Kingdom}
\affiliation{$^4$ Universit\`{a} degli Studi di Palermo, Dipartimento di Energia, Ingegneria dell'Informazione e Modelli Matematici, Viale delle Scienze, Edificio 9, I-90128 Palermo, Italia}
\affiliation{$^5$ Center for Nonlinear Science and Department of Physics, Ningbo University, Ningbo, Zhejiang 315211, China}

\begin{abstract}
We consider the time-dependent resonance interaction energy between two identical atoms, one in the ground state and the other in an excited state, and interacting with the vacuum electromagnetic field, during a nonequilibrium situation such as the dynamical atomic self-dressing process. We suppose the two atoms prepared in a correlated, symmetric or antisymmetric, state. Since the atoms start from a nonequilibrium conditions,  their interaction energy is time dependent.  We obtain, at second order in the atom-field coupling, an analytic expression for the time-dependent resonance interaction energy between the atoms. We show that this interaction vanishes when the two atoms are outside the light-cone of each other, in agreement with relativistic causality, while it instantaneously settles to its stationary value after time $t=R/c$ ($R$ being the interatomic distance), as obtained in a time-independent approach. We also investigate the time-dependent electric energy density in the space around the two correlated atoms, in both cases of antisymmetric (subradiant) and symmetric (superradiant) states, during the dressing process of our two-atom system. We show that the field energy density vanishes in points outside the light-cone of both atoms, thus preserving relativistic causality. On the other hand, inside the light-cone of both atoms, the energy density instantaneously settles to its stationary value. Specifically, for points at equal distance from the two atoms, we find that it vanishes if the two atoms are prepared in the antisymmetric (subradiant) state, while it is enhanced, with respect to the case of atoms in a factorized state, in the symmetric (superradiant) state.  The physical meaning of these results is discussed in detail in terms of interference effects of the field emitted by the two atoms.
\end{abstract}

\maketitle

\section{\label{sec:1}Introduction}
Fluctuation-induced interactions between atoms or molecules, or between macroscopic objects, such as van der Waals and Casimir or Casimir-Polder interactions, have been extensively investigated in the literature, under equilibrium situations \cite{Casimir48, CP48,CT98,CPP95}.   These effects are ubiquitous in a variety of physical, chemical and biological processes \cite{Parsegian06}, playing for example a fundamental role in the stability and  functionality of materials, in biological matter and also in nanotechnologies \cite{WDTRRP16, CC96, CMIC07}.
Several theoretical approaches, based on different physical interpretations, have been proposed in the literature to investigate Casimir and Casimir-Polder interactions, also at finite temperature, highlighting the crucial dependence of these interactions from the system's topology and the magnetodielctric properties of the interacting objects \cite{BKMM09,HIHSPPS09, HS15}.
Although these forces are very tiny (typically, of the order of $10^{-2}$ nN or even less), they have been measured with great accuracy  \cite{SKDL11,DJL11,IKJTDDADL13,MCP09}.
Nevertheless, fundamental problems concerning the dependence of theoretical predictions for real metals, and their accordance with experimental data, are still debated \cite{MBDGFKKLR06,BKMM09,WDTRRP16}.

In recent years, many investigations have been concerned with the resonant forces between atoms or molecules, that occur when one  or more atoms are in an excited state \cite{PT95}. In this case, the interaction can be mediated by the exchange of a real photon, and it can be a very long-range interaction.
If the two atoms are uncorrelated, the resonant Casimir-Polder force is a fourth-order effect in the atom-field coupling, and it scales asymptotically as $R^{-2}$ with the interatomic distance $R$. A different, albeit related, phenomenon takes place when two identical atoms, one excited and the other in the ground state, are prepared in a symmetric or antisymmetric entangled state \cite{CT98, Salam14}. In this case, the excitation is delocalized between the two atoms, and the interaction is a second-order effect, asymptotically scaling as $R^{-1}$. For these reasons, resonance forces are usually much stronger than dispersion interactions.
Resonant and dispersion Casimir-Polder interactions between excited atoms have been object of intense investigations in the past years \cite{RPP04,Berman15, DGL15,MR15,BPRB16}, and they are of considerable importance in a variety of  physical processes. Very recently, the role of resonance interactions in selective interactions between macromolecules in biological systems has been argued \cite{BH85,PP13}.
The possibility of manipulating (enhance or inhibit) the resonance forces, as well as the resonance energy transfer process, through a structured environment (i.e. cavities, waveguides, photonic crystals) has been recently investigated \cite{IFTPRP14, NPR18, FRP18, WPA15, WPGA17, KKY96,KGNH05,WPA16,SK13}.

Different and very interesting effects appear when the overall system is in a non-equilibrium condition, for example when the atoms or molecules are coupled to thermal baths at different temperatures, or are in excited states \cite{CM03, Sherkunov09}. In such cases,  forces that are attractive in equilibrium situations can become repulsive. Moreover, they can be enhanced or suppressed, and their qualitative behaviour (for example the distance dependence) can significantly change. This gives further possibilities of controlling radiative processes through dynamical environments compared to static ones.

Non-equilibrium conditions can be also obtained when some parameter of the system (for example the atomic transition frequency \cite{PV96,PP03,MVP10}, or the field mode frequency in a cavity \cite{BLPRS15}) non-adiabatically changes with time,  bringing the system out of equilibrium. Non-equilibrium conditions are also obtained in the case of moving atoms
\cite{HRS04,MNP14,MHB16,RLMNSZP16}. In all these cases, time evolution of the system toward a new equilibrium configuration is expected. In this context, the dynamical Casimir-Polder interaction between an atom and a reflecting plate, after a non-adiabatic change of some physical parameter of the system, has been investigated \cite{VP08, MVP10, HHSRP14, AVBBRP16,BPRB16}, as well as the time-dependent Casimir-Polder interactions between two or more atoms, during the self-dressing of the atoms \cite{RPP04,PPR07}

The interest in non-equilibrium conditions is both theoretical and experimental. In many realistic physical situations, systems are in a nonequilibrium condition (for example, the temperatures of the interacting objects can be different from the environment's temperature or from each other, or the objects involved can be in relative motion). On the other hand, it is also a fundamental issue to investigate whether and how  nonequilibrium initial conditions can influence the dynamics of a given system, or modify radiation-mediated (resonance or dispersion) interactions between atoms or molecules during their dynamic evolution. An important conceptual issue is also investigating how the new equilibrium condition is approached by the system \cite{AVBBRP16}.

In this paper, we investigate the time-dependent resonance interaction energy between two identical two-level atoms, during the dynamical self-dressing process of the system. We consider the two atoms in free space, and interacting with the electromagnetic field in its vacuum state.
We  suppose that our atomic system, with one atom in the excited state and the other in the ground-state, is initially prepared in a correlated (symmetric or antisymmetric) state. The symmetric (antisymmetric) state is the well-known superradiant (subradiant) state in the Dicke theory \cite{Dicke}. Dynamical properties of these states have been extensively explored in the literature, also in connection with the
problem of causality in quantum electrodynamics \cite{BCPPP90}.
Since the initial state of the system is not an eigenstate of the interacting Hamiltonian, it evolves in time, and we investigate the dynamical resonance interaction during the dressing process of the two-atoms-field interacting system. The interaction energy is therefore time dependent, because of the dynamical self-dressing of the atoms, that start from a non-equilibrium state. Using perturbation theory up to second-order, we obtain the time-dependent expression for the resonance interaction energy between the two atoms. We find that this interaction is equal to zero when the two atoms are outside the light-cone of each other, in agreement with relativistic causality, while it sharply settles to its stationary value after time $t=R/c$ ($R$ being the interatomic distance). Thus, the system approaches its equilibrium configuration immediately after the causality time.
We investigate in detail the contributions from virtual and real photons emitted during the dynamical dressing of the system, and show that the virtual photons contribution (due to the counterrotating terms of the Hamiltonian) is essential to ensure the causality in the time-dependent resonance interaction energy.  We show that the behaviour of the time-dependent resonance interaction can be related to an interference effect between the contributions from rotating and counterrotating terms, bringing the system to its local equilibrium configuration immediately after the causality time. We finally investigate the time-dependent electric energy density in the space around the two correlated atoms and discuss its behaviour  in both cases of subradiant and superradiant initial states. In particular, we show that for points at the same distance from the two atoms, the value of the electric energy density around the two-atom system is zero if the atoms are prepared in the subradiant state, while for atoms in the superradiant superposition it is doubled with respect to the case of uncorrelated atoms. Finally, we discuss the possibility to probe this behavior of the electric energy density through a measurement of the Casimir-Polder force on a third atom, placed nearby the two-atom system.

This paper is structured as follows. In section \ref{sec:2} we introduce our model and analyse the time-dependent resonance interaction energy between the two atoms, both in the case of superradiant and subradiant states. In section \ref{sec:3} we investigate the time-dependent electric energy density surrounding the atoms. Finally, section \ref{sec:4} is devoted to final remarks and conclusions.

\section{\label{sec:2} Time-dependent resonance interaction energy}

We consider two identical atoms $A$ and $B$, respectively placed at positions $\br_A$ and $\br_B$, and interacting with the quantum electromagnetic field in the vacuum state. We model the atoms as two-level systems.
The Hamiltonian for the system in the multipolar coupling scheme in the Coulomb gauge and within the dipole approximation, is \cite{CT98,CPP95}
\begin{eqnarray}
\label{eq:0}
&\ & H=H_0+H_I ,
\end{eqnarray}
where
\begin{eqnarray}
\label{eq:1a}
H_0=\hbar\wo (S^A_z+S^B_z)+\sum_{\mathbf{k}j}\hbar\wk \akjd\akj
\end{eqnarray}
is the free Hamiltonian, and
\begin{eqnarray}
\label{eq:1}
H_I=&-&{\boldsymbol\mu_{A}}\cdot\mathbf{E}(\br_A)-\boldsymbol\mu_{B}\cdot\mathbf{E}(\br_B)
\end{eqnarray}
is the interaction Hamiltonian.
In the expressions above, 
$S_z={\frac 1 2}(\vert e\rangle\langle e\vert-\vert g\rangle\langle g\vert)$ is the pseudospin atomic operator, $\vert g\rangle$  and $\vert e\rangle$ are the ground and the excited atomic states  with energies $\mp \hbar \omega_0/2$, respectively  ($\omega_0$ being the transition frequency of both atoms), and $\bmu=e\br = \bmu^{eg}(S_+ + S_-)$ is the atomic dipole moment operator, with its matrix elements $\bmu^{eg}=\langle e \mid \bmu \mid g \rangle$ assumed real. Also, $\akj$ ($\akjd$) is the bosonic annihilation (creation) operator for photons with wavevector $\mathbf{k}$ and polarization $j$. Finally, $\mathbf{E}(\br, t)$ is the transverse displacement field operator, that outside the atoms coincides with the total (longitudinal plus transverse) electric field,

\begin{equation}
\label{eq:2}
{\bf E}({\bf r},t)=i\skj\sqrt{\frac{2\pi\hbar \wk}{V}}\ekj \left( \akj(t)\ekr - \akjd(t)\ekrm \right)
\end{equation}
($\ekj$ are the polarisation unit vectors of the electromagnetic field, assumed real). In our case of two-level atoms, the interaction Hamiltonian in (\ref{eq:1}) can be conveniently rewritten as
\begin{eqnarray}
\label{eq:3}
&\ & H_{I} (t)=-i\sAB \skj\sqrt{\frac{2\pi\hbar ck}{V}}(\ekj\cdot\bmu_i^{eg})\nonumber\\
&\times & \! \! \! \!\left[\akj \ekri\left(S_{+}^{i}+\lambda S_{-}^{i} \right)-\akjd\ekrmi\left(S_{-}^{i}+\lambda S_{+}^{i} \right)\right] ,
\end{eqnarray}
where $\lambda = 0,1$ is a parameter that allows to easily identify the role of counterrotating terms: $\lambda = 1$ gives the full Hamiltonian, while for $\lambda =0$ we get the Hamiltonian in the rotating wave approximation (RWA).

We assume that the two atoms are initially prepared in one of the two entangled superradiant or subradiant Dicke states \cite{Dicke}
\begin{eqnarray}
\label{eq:4}
\vert \phi_{\pm}\rangle=\frac{1}{\sqrt{2}}\big( \vert e_A, g_B\rangle\pm \vert g_A, e_B\rangle\big) ,
\end{eqnarray}
and that the field is in the vacuum state $\vac$.
The excitation is therefore delocalized between the two atoms.
The initial state of the system is thus the bare state
\begin{equation}\label{eq:4a}
\vert \psi_{\pm}\rangle=\vert \phi_{\pm}\rangle\vac ,
\end{equation}
with {\em bare} (noninteracting) energy $E_{\psi}=0$. Because of the atom-field interaction, this state is not an eigenstate of the total Hamiltonian, and therefore it will evolve in time according to
\begin{equation}
\vert\psi_{\pm}(t)\rangle=e^{-i H t/\hbar}\vert\psi_{\pm}\rangle ,
\end{equation}
where $H$ is the total Hamiltonian \eqref{eq:1} of the system.
We wish to investigate the time-dependent energy shift of the two-atom system, during the dynamical self-dressing process of the system.
The physical situation we are considering is equivalent to switching on the atom-field coupling at $t=0$; after that, the system is not longer in an equilibrium situation and it will unitarily evolve starting from its bare state $\vert\psi_{\pm}\rangle$.

In order to obtain the time-dependent interaction energy, we follow the same approach used in \cite{VP08, MVP10, AVBBRP16} for the dynamical atom-surface Casimir-Polder interaction. Specifically, we first write down the Heisenberg equations for field and atomic operators, and solve them iteratively at the lowest significant order; the time dependent energy shift of the system is then obtained by evaluating the quantity
\begin{eqnarray}
\label{eq:5}
\Delta E(t)=\frac{1}{2}\langle \psi_{\pm}\vert H_I^{(2)}(t)\vert\psi_{\pm}\rangle ,
\end{eqnarray}
where $H_I^{(2)}(t)$ is the interaction Hamiltonian \eqref{eq:3} in the Heisenberg representation at second order in the coupling (obtained by substitution of the second-order solution of the Heisenberg equations for atom and field operators).
This method is a direct generalization to time-dependent situations of the following (time-independent) relation, as obtained by second-order stationary perturbation theory \cite{CPP83},
\begin{eqnarray}
\label{eq:6}
\Delta E=\frac{1}{2}\langle \psi_D\vert H_I\vert\psi_D\rangle ,
\end{eqnarray}
where $\vert \psi_D\rangle$ is the second-order interacting (dressed) state of the system, and $H_I$ is the interaction Hamiltonian in the Schr\"{o}dinger representation.

It should be observed that, since the time evolution is unitary, the total energy of the system is conserved during its self-dressing process. In particular, this means that the average value of the total Hamiltonian $H^{(2)}(t)$ (Heisenberg representation and evaluated at second order in the coupling) on the initial bare state $\vert\psi_{\pm}\rangle$ does not depend on time, that is
\begin{eqnarray}
\label{eq:6a}
&\ &\langle \psi_{\pm}\vert H^{(2)}(t)\vert\psi_{\pm}\rangle= \langle \psi_{\pm}\vert H^{(2)}_{0}(t)\vert\psi_{\pm}\rangle\nonumber\\
&\ & +\langle \psi_{\pm}\vert H^{(2)}_I(t)\vert\psi_{\pm}\rangle=0 ,
\end{eqnarray}
for any $t$, where $H^{(2)}(t)$ is the total Hamiltonian in the Heisenberg representation, evaluated at the second-order in the coupling. Eq. \eqref{eq:6a} can be easily verified by evaluating, at the second-order in the coupling, the single contributions to the total energy of the system, that is the average value of  atomic, field and interaction Hamiltonians on the initial bare state of the system, similarly to the case of the atom-plate dynamical Casimir-Polder interaction energy \cite{AKBRP17}. This shows that the time evolution is unitary at the order considered.
Nevertheless, we will see that the average value of the interaction Hamiltonian $H_I^{(2)}(t)$ on the bare state $\vert\psi_{\pm}\rangle$, related to the local interaction energy of the atoms with the field evaluated at their position, is nonvanishing and depends on time; actually, it gives the time-dependent resonance interaction energy of our two-atom system.

As mentioned above, we first obtain the expressions of field and atomic operators in the Heisenberg representation, at the lowest significant order. After some algebra, we obtain
\begin{widetext}
\begin{eqnarray}
\label{eq:7}
\akj(t)&=&\ewkm \akj (0)+\sAB e^{-i\bk\cdot\br_i}\ewkm\sqrt{\frac{2\pi\wk}{\hbar V}}[\bmu_i^{eg}\cdot\ekj][S_{+}^i(0)F(\wo+\wk,t)+\lambda S_{-}^i(0)F(\wk-\wo,t)], \\
\label{eq:8}
S_{-}^i(t)&=&e^{-i\wo t}S_{-}^i(0)+2S_z^i(0)e^{-i\wo t}\skj\sqrt{\frac{2\pi\wk}{\hbar V}}[\bmu_i^{eg}\cdot\ekj][\akj (0)e^{i\bk\cdot\br_i}F(\wo-\wk,t)-\lambda\akjd (0)e^{-i\bk\cdot\br_i}F(\wk+\wo,t)]\, ,\nonumber\\
\end{eqnarray}
\end{widetext}
$(i=A,B)$ where we have defined the function
\begin{equation}
\label{eq:9}
F(x,t)=\frac{e^{ix t}-1}{i x}\, .
\end{equation}

Now, using Eqs. \eqref{eq:7} and \eqref{eq:8} in the expression of  $H^{(2)}_{I}(t)$, as given in \eqref{eq:3}, and taking only terms up to the second order in the coupling, we obtain the explicit expression of $H^{(2)}_{I}(t)$
\begin{widetext}
\begin{eqnarray}
\label{eq:10}
H_{I}^{(2)}(t)=&\ &-\frac{2\pi i}{V}\sqAB\ssAB \skj \wk (\ekj\cdot\bmu_{q}^{eg}) (\ekj\cdot\bmu_{s}^{eg})e^{i\bk\cdot(\br_q-\br_s)}\ewkm\left( S_+^q(0)\ewo+\lambda S_-^q(0)\ewom \right)
\nonumber\\
&\ &\times \left(S_-^s(0)F(\wk-\wo,t) +\lambda S_+^s(0)F(\wk+\wo,t)\right)
\nonumber\\
&\ &-\frac{4\pi i}{V} \sqAB\skj\skjp\sqrt{\wk \wkp}(\ekj\cdot \bmu_{q}^{eg})(\ekjp\cdot\bmu_{q}^{ge})S_{z}^{q}(0)\akj(0)e^{i(\bk\cdot\br_q- \wk t)}\nonumber\\
&\ &\times \left[ \akjdp (0) e^{-i\bkp\cdot\br_q} \left( F^*(\wo-\wkp,t) \ewo
-\lambda^2 F(\wo+\wkp,t) \ewom \right) -\lambda \akjp (0) e^{i\bkp\cdot\br_q} \right.
\nonumber \\
&\ &\times  \left( F^*(\wo+\wkp,t) \ewo - F(\wo-\wkp,t) \ewom \right) \Big] +\mbox{h.c.} ,
\end{eqnarray}
\end{widetext}

We now evaluate the average value of \eqref{eq:10} on each of the two correlated states $\vert \psi_{\pm}\rangle$.  We take into account only terms depending on the interatomic distance $\bR=\br_A-\br_B$, which are the only ones relevant for the dynamical interatomic interaction energy, from which in a quasi-static approach the resonance force can be obtained by taking the negative derivative with respect to the interatomic distance; the other terms just give single-atom energy shifts. After some algebra, we obtain the time-dependent resonance interaction energy
\begin{eqnarray}
\label{eq:11}
&\ &\Delta E(t)=\frac{1}{2}\langle\psi_{\pm}\vert H_I^{(2)}(t)\vert\psi_{\pm}\rangle\nonumber\\
&\ & =\mp i\frac{\pi c}{2V}\sum_{q,s=A,B (q\neq s)} \skj k (\ekj\cdot\bmu_q^{eg}) (\ekj\cdot\bmu_s^{ge})
\nonumber\\
&\ &\times\Biggl [e^{i\bk\cdot(\br_q-\br_s)}\left(F^*(\wk-\wo,t) +\lambda^2 F^*(\wk+\wo,t)\right)\Biggr]\nonumber\\
&\ &+\mbox{c.c.} ,
\end{eqnarray}
where the upper (lower) sign respectively refers to the symmetric (antisymmetric) state, and in the summation over $q,s$ we must take $q\not=s$.  In the continuum limit, using $\sum_{\bk}\rightarrow V/(2\pi)^3 d^3{\bf k}$ and the polarization sum
$\sum_j (\ekj)_{\ell}(\ekj)_m=\delta_{\ell m}-{\bf\hat{k}}_{\ell}{\bf\hat{k}}_{m}$ ($\ell, m=x,y,x$), the angular integration yields
\begin{eqnarray}
\label{eq:12}
&\ & \int d\Omega_{\bk} (\delta_{\ell m}-{\bf\hat{k}}_{\ell}{\bf\hat{k}}_{m})e^{\pm i\bk\cdot \bR}\nonumber\\
&\ & = \frac{4\pi}{k^3}(-\delta_{\ell m}\nabla^2+\nabla_{\ell}\nabla_m) \frac{\sin(k R)}{R} ,
\end{eqnarray}
(the differential operators act on $R=\mid \bR \mid$), and we obtain
\begin{eqnarray}
\label{eq:13}
&\ &\Delta E(t)=\mp\frac{c}{\pi}(\bmu^{eg}_A)_{\ell}(\bmu^{eg}_B)_m
F_{\ell m}^R \frac{1}{R}\int_{0}^{\infty} \! \! dk\sin(kR)\nonumber\\
&\ & \times\left(\frac{1-\cos(ck-\wo)t}{ck-\wo}+\lambda^2\frac{1-\cos(ck+\wo)t}{ck+\wo}\right) ,
\end{eqnarray}
where we have defined the differential operator $F_{\ell m}^R=(-\delta_{\ell m}\nabla^2+\nabla_{\ell}\nabla_{m})^R$ acting on the variable $R$. The Einstein convention for repeated indices has been used.

In the expression \eqref{eq:13}, we can distinguish a time-independent term, that is the same obtained in the static case \cite{IFTPRP14,NPR18}, and a term explicitly depending on time, related to the time-dependent self-dressing of the entangled two-atom system.
We note that, in contrast to the stationary case where a pole at $ck=\wo$ appears (related to the emission of a real photon from the excited atom), in the present case there are no poles in \eqref{eq:13}; therefore, there is not ambiguity in circumventing the pole in the frequency integration. We should however note that perturbative approach is valid only for times much smaller than the lifetime of the excited atom.
This observation is relevant also in connection with controversial results in the literature concerning with the long-distance behaviour (with space oscillations or not) of the dispersion Casimir-Polder interaction between an excited- and a ground-state atom \cite{RPP04,MR15,Berman15}.

To obtain the explicit expression of the dynamical resonance interaction, we now evaluate the integrals in \eqref{eq:13}.
Since the time-dependent integrals diverge on the light-cone, $R=ct$, we perform the calculation by considering separately the two time regions $t<R/c$ and $t>R/c$ (i.e. before and after the causality time).  As discussed in \cite{VP08,AVBBRP16}, these divergences are related to the assumption of pointlike field sources, as well as to our bare-state initial condition \cite{PV96,PP03,MVP10}. Similar divergences appear during the time-dependent self-dressing of a single initially bare source \cite{BCPPP90}, as well as in the field energy densities nearby a reflecting mirror \cite{FS98,BP12,BBLPRS15} or a point-like field source \cite{PRS13,CPP88}.
For $t<R/c$, we obtain
\begin{widetext}
\begin{eqnarray}
\label{eq:14}
\Delta E(t)=&\pm&\frac{1-\lambda^2}{2\pi}(\bmu^{eg}_A)_{\ell}(\bmu^{eg}_B)_m
F_{\ell m}^R \frac{1}{R}\left[\sin(k_0 R)\biggl(2\text{Ci}[k_0 R] -\text{Ci}[k_0(R - ct)] -\text{Ci}[k_0(R + ct)]\biggr)\right.\nonumber\\
 &-& \left. \cos(k_0 R) \biggl(2\text{si}[k_0R]-\text{si}[k_0(R-ct)]-\text{si}[k_0(R+ct)]\biggr)\right] ,
\end{eqnarray}
while for $t>R/c$ we have
\begin{eqnarray}
\label{eq:15}
\Delta E(t)=&\mp&(\bmu^{eg}_A)_{\ell}(\bmu^{eg}_B)_m
F_{\ell m}^R \frac{1}{R} \cos(k_0R)\nonumber\\
&\pm&\frac{1-\lambda^2}{2\pi}(\bmu^{eg}_A)_{\ell}(\bmu^{eg}_B)_m
F_{\ell m}^R \frac{1}{R}\left[\sin(k_0 R)\biggl(2\text{Ci}[k_0 R] -\text{Ci}[k_0(ct-R)] -\text{Ci}[k_0(ct+R)]\biggr)\right.\nonumber\\
&-&\left. \cos(k_0R) \biggl(2\text{si}[k_0R]+\text{si}[k_0(ct-R)]-\text{si}[k_0(ct+R)]\biggr)\right] ,
\end{eqnarray}
\end{widetext}
where the cosine and sine integral functions,  $\text{Ci}(x)$ and  $\text{si}(x)=\text{Si}(x)-\pi/2$, have been introduced \cite{GR00,AS}.
Equations \eqref{eq:14} and \eqref{eq:15} give the dynamical resonance interaction energy as a function of $t$, during the self-dressing process of the two-atom system. The upper (lower) sign respectively refers to  symmetric (antisymmetric) state, yielding correlation (anticorrelation) between the atomic dipole moments at $t=0$, being
\begin{equation}
\langle\psi_{\pm}\vert (\bmu_{A})_i(\bmu_{B})_j\vert\psi_{\pm}\rangle=\pm(\bmu^{eg}_{A})_i(\bmu^{ge}_{B})_j
\end{equation}
(on the contrary, in fourth-order dispersion forces the atomic dipoles are correlated by vacuum fluctuations \cite{PT93,PPR03,CCPPP07}).

For $\lambda=0$ (RWA), only the rotating terms of the Hamiltonian contribute to the dynamical resonance interaction, and they give a non vanishing time-dependent interaction energy, even before the causality time $t=R/c$.  On the contrary, if we also include the contribution from virtual photons given by the counterrotating terms in the Hamiltonian (that is, if we set $\lambda=1$), we immediately obtain
\begin{eqnarray}
\label{eq:16}
\Delta E(t)=\mp(\bmu_A)_{\ell}(\bmu_B)_{m}F_{\ell m}^R\frac{\cos(k_0 R)}{R}\theta(ct-R) ,
\end{eqnarray}
and the causal behaviour is fully recovered. This clearly shows that  the contribution of virtual photons given by the counterrotating terms is essential to ensure the causal behaviour of the dynamical resonance interaction energy, similarly to the case of two ground-state atoms \cite{BCPPP90} or in the Fermi problem \cite{CPPP95}.

Performing the derivatives with respect to $R$ in \eqref{eq:16}, we get
\begin{eqnarray}
\label{eq:17}
&\ &\Delta E(t)=\pm(\bmu_A)_{\ell}(\bmu_B)_{m}V_{\ell m}(k_0,R)\theta(ct-R),
\end{eqnarray}
where $V_{\ell m}(k_0, R)$ is the potential tensor \cite{CT98}
\begin{widetext}
\begin{eqnarray}
\label{eq:18}
V_{\ell m}(k_0,R)=\frac{1}{R^3}\biggl[ (\delta_{\ell m}-3{\hat{R}}_{\ell}{\hat{R}}_m)\left (\cos (k_0R)+k_0R\sin (k_0R) \right)-(\delta_{\ell m}-{\hat{R}}_{\ell}{\hat{R}}_m)k_0^2R^2\cos (k_0R)\biggr].
\end{eqnarray}
\end{widetext}

Thus, there is not interaction between the two atoms before time $t=R/c$. After this time, the dynamical resonance interaction instantaneously settles to its stationary value. An observer measuring the resonance force on one atom (for example atom $A$), will detect the full value of the stationary force immediately after the causality time $t=R/c$.
This behavior is very different from that of the dynamical Casimir-Polder interaction between two uncorrelated atoms (which is a fourth-order effect) \cite{RPP04}, or the dynamical atom-plate Casimir-Polder energy between an atom and a conducting plate (which is a second-order effect) \cite{VP08, MVP10}; in fact, in the latter cases the time-dependent interaction energy settles to its stationary value only for times larger than $\wo^{-1}$, which is the time scale of the dynamical atomic dressing. Otherwise, in the present case the time-dependent contributions related to the dynamical dressing process of the two atoms cancel with each other (at least for what concerns the interaction energy), and the system approaches its local equilibrium configuration immediately after the causality time $R/c$.

The sharp change of the time-dependent resonance interaction energy \eqref{eq:17} at $t=R/c$ is related to the assumption of point-like atoms inherent in the dipole approximation and to the specific bare initial state considered (two correlated atoms, one excited and the other in the ground-state, and the field in the vacuum state). In our initial state, correlations between the two atoms are present, leading to a cancellation of the time-dependent contributions to the dynamical self-dressing of the system. Thus the peculiar behaviour  of the interaction at the causality time $t=R/c$, can be ascribed to an interference effect between the contributions of real and virtual processes occurring during the time-dependent self-dressing of the two-atom system. Also, we have a bare  initial state, so the two atoms can feel each other only after the causality time $t=R/c$, and this explains why the interaction energy vanishes for $t<R/c$. On the other hand, in the multipolar coupling scheme the interaction between the atoms is mediated by the transverse displacement field that, outside the atoms, coincides with the total electric field, thus obeying a fully retarded wave equation with point-like sources. This yields retarded solutions at any order of the coupling, and we thus expect that this sharp behaviour should not be limited to our second-order solution.

\begin{figure}[!htbp]
\centering
\includegraphics[scale=0.9]{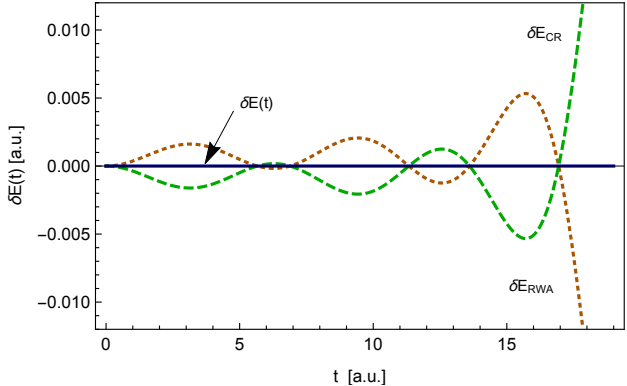}
\caption{Plots of the time-dependent contributions to the dynamical resonance interaction energy as a function of time, for $t<R/c$. Energy and time are in arbitrary units. The interatomic distance is set to $R=20$, while $c=1$ and $k_0=1$; thus $t<R/c$ means $t<20$. The orange dotted line and the green dashed line represent, respectively, the time-dependent contributions from rotating ($\delta E_{RWA}(t)$) and counterrotating  ($\delta E_{CR}(t)$) terms. The blue solid line is the total dynamical resonance interaction energy, before the causality time. This plot points out that the contributions $\delta E_{\it RWA}$ and $\delta E_{\it CR}$ oscillate in time out of phase, and that their sum vanishes at all times $t<R/c$. Both contributions diverge on the light cone $t=R/c$.}
\label{fig:1}
\end{figure}

\begin{figure}[!htbp]
\centering
\includegraphics[scale=0.9]{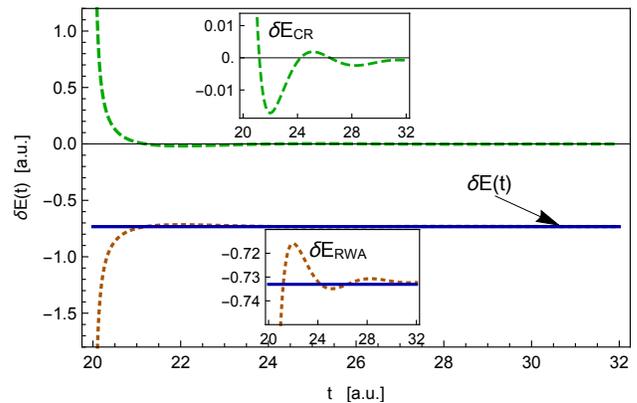}
\caption{Plots of the time-dependent contributions, $\delta E_{\it RWA}(t)$ and $\delta E_{\it CR}(t)$, to the dynamical resonance interaction energy, for $t>R/c$. Energy and time are in arbitrary units. Parameters are  $R=20$, $c=1$, and $k_0=1$; thus $t$ is larger than $20$. The orange dotted line and the green dashed line represent the time-dependent contributions from rotating ($\delta E_{RWA}(t)$) and counterrotating  ($\delta E_{CR}(t)$) terms, respectively. The solid blue line is the dynamical resonance interaction energy, after the causality time. The plots show that both contributions from real and virtual photons oscillate in time, and that their sum, immediately after $t=R/c$, is time-independent and coincides with the stationary value of the resonance interaction energy, as obtained by a time-independent approach. The two inserts are enlargements showing the presence of time oscillations of both contributions soon after the causality time.}
\label{fig:2}
\end{figure}

This conclusion can be mathematically understood by considering the integrals in \eqref{eq:13}.
For $\lambda=1$, evaluation of the two integrals in \eqref{eq:13} can be easily performed by considering separately time-independent and time-dependent terms.
For the time-independent terms, we have
\begin{eqnarray}
\label{eq:19}
&\ &I_1=\mathcal{P}\int_{0}^{\infty}dk\sin(kR)\left(\frac{1}{ck-\wo}+\frac{1}{ck+\wo}\right)\nonumber\\
&\ &=\mathcal{P}\int_{-\infty}^{\infty}dk\frac{\sin(kR)}{ck+\wo},
\end{eqnarray}
where $\mathcal{P}$ indicated the principal value and  in the first integral we performed the variable change $k\rightarrow -k$.
 This integral has a pole in $k=-k_0=-\wo /c$ and, using the residue theorem, we obtain
 \begin{eqnarray}
 \label{eq:20}
 I_1=\frac{\pi}{c}\cos(k_0 R) .
\end{eqnarray}
The time-dependent integral can be similarly evaluated, obtaining
\begin{eqnarray}
\label{eq:21}
&\ & I_2(t)=\mathcal{P}\int_{0}^{\infty}dk\sin(kR)\nonumber\\
&\ &\times\left(\frac{\cos[(ck-\wo)t]}{ck-\wo}+\frac{\cos[(ck+\wo)t]}{ck+\wo}\right)\nonumber\\
&\ &=\mathcal{P}\int_{-\infty}^{\infty}dk\frac{\sin(kR)}{ck+\wo}\cos[(ck+\wo)t]\nonumber\\
&\ &=\frac{\pi}{c}\cos(k_0 R)\theta(R-ct).
\end{eqnarray}
This clearly shows that the time-dependent resonance interaction energy, depending on $I_1-I_2$, vanishes for $t<R/c$, while it is nonvanishing and  time-independent for $t>R/c$. The system thus sharply approaches its equilibrium configuration at the causality time $R/c$. For $t>R/c$, the dynamical energy shift of the two-atom system \eqref{eq:17} coincides with the stationary value obtained for the time-independent fully dressed state \cite{IFTPRP14}, that is
\begin{equation}
\label{eq:22}
\Delta E(t)=\frac{1}{2}\langle \psi_{\pm}\vert H_I^{(2)}(t)\vert\psi_{\pm}\rangle
\rightarrow \Delta E = \langle \psi_{D}\vert H\vert\psi_{D}\rangle ,
\end{equation}
$\vert\psi_{D}\rangle$ being the dressed state of the atom-field system and $H$ the total Hamiltonian.
This behaviour is clearly illustrated in figures \ref{fig:1} and \ref{fig:2}, that show the contributions to the dynamical resonance interaction energy from the rotating and counterrotating terms, in the two time intervals, $t<R/c$ and $t>R/c$, as a function of $t$. Fig. \ref{fig:1} shows that for $t<R/c$  the two contributions oscillate in time out of phase, and diverge on the light cone $t=R/c$. As mentioned, their sum vanishes, coherently with relativistic causality. Fig. \ref{fig:2} shows that, after the causality time, time-dependent contributions cancel out, and the resonance-interaction static value is recovered.

This finding suggests that the peculiar sharp behaviour of the dynamical resonance interaction is due to an interference effect between the virtual-photon and the real-photon contributions; these photons are generated by the two identical atoms during their dynamical self-dressing process.

As mentioned, our results are valid for times shorter than the decay time of the excited state (of the order of $10^{-8}$ s). For typical values of the atomic parameters $\mu\simeq 10^{-29}$ C m,  $k_0\simeq 10^{7}$ m$^{-1}$, and for an interatomic distance of $10^{-6}$ m, the resonance force between the two atoms is of the order of $10^{-21}$ N, which is several orders of magnitude larger than the dispersion Casimir-Polder interaction between atoms, that can be measured directly \cite{Beguin13}, and comparable with the atom-surface Casimir-Polder interactions for which many measurements exist \cite{BKMM09}. However, the resonance interaction energy, and the consequent interatomic force, is very difficult to observe directly, since it requires the system be prepared and mantained in an entangled state for a sufficiently long time. The correlated state is very fragile, and the coherent superposition can be easily destroyed by spontaneous emission. Very recently, the possibility to control decoherence effects through a structured environment, such as a photonic crystal, has been explored \cite{IFTPRP14, NPR18}, thus suggesting possible experimental setups to observe such interaction for atoms placed inside a structured environment.

In order to gain further physical evidence of the processes involved during the dynamical dressing, in the next section we shall investigate the time-dependent electric field energy density during the self-dressing of the two-atom system.

\section{\label{sec:3} Time-dependent electric energy density around the two entangled atoms.}

In order to further investigate the time-dependent self-dressing process of the two entangled atoms and its relation with the resonance interaction energy, we now evaluate the (time-dependent) electric energy density nearby the two-atom system, for both symmetric and antisymmetric states. It is given by
\begin{eqnarray}
\label{eq:23}
&\ &\langle \psi_{\pm}\vert \mathcal{H}_{el}(\br,t)\vert\psi_{\pm}\rangle=\frac{1}{8\pi}\,\, \langle \psi_{\pm}\vert E^{2}(\br, t)\vert\psi_{\pm}\rangle
\end{eqnarray}
(the electric field operator is in the Heisenberg representation). Up to second-order in the coupling, we have
\begin{eqnarray}
\label{eq:24}
&\ &\frac{1}{8\pi} E^{2}(\br, t)=\frac{1}{8\pi} \left( {\bf E}^{(0)}(\br, t)\cdot {\bf E}^{(0)}(\br, t) \right.\nonumber\\
&\ &\left. + {\bf E}^{(1)}(\br, t)\cdot {\bf E}^{(1)}(\br, t)+{\bf E}^{(2)}(\br, t)\cdot {\bf E}^{(0)}(\br, t)\right.\nonumber \\
&\ &\left.+{\bf E}^{(0)}(\br, t)\cdot {\bf E}^{(2)}(\br, t)\right) ,
\end{eqnarray}
where the superscript indicates the perturbative order.
Iterative solution of the Heisenberg equations for the field annihilation operators at the leading order, gives
\begin{widetext}
\begin{eqnarray}
\label{eq:25}
&\ &\akj^{(0)}(t)=\akj(0)\ewkm ,\\
\label{eq:26}
&\ &\akj^{(1)}(t)=\left(\frac{2\pi \wk}{V\hbar}\right)^{1/2}\sqAB\left[(\bmu_q^{eg}\cdot\ekj)e^{-i\bk\cdot\br_q}\ewkm\left(S_{+}^q(0)F(\wo+\wk,t)+S_{-}^q(0)F(\wk-\wo,t)\right)\right]  ,\\
\label{eq:27}
&\ &a_{\bk j}^{(2)}(t)=-\frac{4\pi}{\hbar V}\sqAB S_z^q(0) e^{-i(\bk\cdot \br_q-\wk t)}\skjp\sqrt{\wk\wkp}(\bmu_q^{eg}\cdot\ekj)(\bmu_q^{ge}\cdot\ekjp)\nonumber\\
&\ &\times i\left\{\akjp(0)e^{\bkp\cdot\br_q}\left[i F(\wk-\wkp,t)\left(\frac{1}{\wo+\wk}+\frac{1}{\wo-\wkp}\right)
 -\frac{1}{\wo+\wkp}F(\wo+\wk,t)-\frac{1}{\wo-\wkp}F^*(\wo-\wk,t)\right]\right.\nonumber\\
&\ &\left. -\akjdp(0)e^{-i\bkp\cdot\br_q}\left[ F(\wk+\wkp,t)\left(\frac{1}{\wo-\wkp}+\frac{1}{\wo+\wkp}\right) -\frac{1}{\wo-\wkp}F(\wo+\wk,t)-\frac{1}{\wo+\wkp}F^*(\wo-\wk,t)\right]\right\} .
\end{eqnarray}
\end{widetext}
Since second-order field operators depend on $S_z^A(0)$ and $S_z^B(0)$, as expressed by \eqref{eq:27}, the contribution $\langle \psi_{\pm}\vert {\bf E}^{(2)}(\br, t)\cdot {\bf E}^{(0)}(\br, t)\vert\psi_{\pm}\rangle$ in \eqref{eq:24} vanishes, being $\langle \psi_{\pm}\vert S_z^{A,B}(0)\vert\psi_{\pm}\rangle=0$. The only nonvanishing term comes from
$\langle\psi_{\pm}\vert {\bf E}^{(1)}(\br, t)\cdot {\bf E}^{(1)}(\br, t)\vert\psi_{\pm}\rangle$ and,
to evaluate it, we first substitute (\ref{eq:26}) into the expression of the electric field (\ref{eq:2}), obtaining
\begin{widetext}
\begin{eqnarray}
\label{eq:28}
&\ &{\bf E}^{(1)}(\br,t) = i\frac{2\pi}{V}\sqAB\skj \wk\ekj(\bmu_q^{eg}\cdot\ekj)\left[e^{i\bk\cdot(\br-\br_q)}\ewkm\left(S_{+}^q(0)F(\wo+\wk,t)+S_{-}^q(0)F(\wk-\wo,t)\right)\right]+\, \mbox{h.c.} .\nonumber\\
\end{eqnarray}
\end{widetext}

The time-dependent electric energy density is then obtained  by substitution of (\ref{eq:28}) into (\ref{eq:24}) (we disregard the bare space-uniform vacuum field contributions coming from the zeroth-order term).  After some algebra, involving polarization sum and integration over $\bk$, we finally obtain
\begin{widetext}
\begin{eqnarray}
\label{eq:29}
&\ &\langle \psi_{\pm}\vert \mathcal{H}_{el}\vert\psi_{\pm}\rangle = \langle \psi_{\pm}\vert \mathcal{H}_{el}^{(A)}\vert\psi_{\pm}\rangle
+ \langle\psi_{\pm}\vert \mathcal{H}_{el}^{(B)}\vert \psi_{\pm}\rangle
+\,\langle \psi_{\pm}\vert \mathcal{H}_{el}^{(AB)}\vert\psi_{\pm}\rangle ,
\end{eqnarray}
where
\begin{eqnarray}
\label{eq:30a}
&\ &\langle \psi_{\pm}\vert \mathcal{H}_{el}^{(A)}\vert\psi_{\pm}\rangle=\frac{1}{8\pi}\Re\left((\bmu_A^{eg})_m (\bmu_A^{ge})_n F_{\ell m}^{R_A}\frac{e^{i k_0 R_A}}{R_{A}} F_{\ell n}^{R_A}\frac{e^{- i k_0 R_A}}{R_{A}} \right)\theta(ct-R_A) ,\\
\label{eq:31a}
&\ &\langle \psi_{\pm}\vert \mathcal{H}_{el}^{(B)}\vert\psi_{\pm}\rangle= \langle \psi_{\pm}\vert \mathcal{H}_{el}^{(A)}\vert\psi_{\pm}\rangle \,\, \mbox{with}\,\,\,  A\leftrightarrows B , \\
\label{eq:32a}
&\ &\langle \psi_{\pm}\vert \mathcal{H}_{el}^{(AB)}\vert\psi_{\pm}\rangle=\pm \frac{1}{8\pi}\Re\left((\bmu_A^{eg})_m (\bmu_B^{ge})_n F_{\ell m}^{R_A}\frac{e^{i k_0 R_A}}{R_{A}} F_{\ell n}^{R_B}\frac{e^{-i k_0 R_B}}{R_B} +\mbox{c.c}\right)\theta(ct-R_A)\theta(ct-R_B) ,
\end{eqnarray}
\end{widetext}
where ${\bf R}_{A(B)}=\br-{\bf r}_{A (B)}$ is the distance between the observation point $\br$ (where the electric energy density is evaluated) and atom $A (B)$. The $ +/- $ sign refers to the superradiant/subradiant state given in Eq. \eqref{eq:3}.
Eq. (\ref{eq:29}) describes the time-dependent electric energy density emitted by atoms $A$ and $B$,  and we may separate two contributions. The first contribution, given by \eqref{eq:30a} and \eqref{eq:31a}, is related to the retarded field emitted by each atom, and it is causal, due to the presence of the Heaviside step function. This contribution vanishes if point $\br$ is outside the causality sphere of both atoms $A$ and $B$, that is if $R_A, R_B > c t$, as expected. On the other hand, the second contribution, Eq. \eqref{eq:32a}, is a sort of {\em interference} term, related to the  electric field radiated by the overall two-atom system. Inspection of \eqref{eq:30a}-\eqref{eq:32a} clearly shows that if the point $\br$ is outside the light-cone of both atoms, the electric energy density \eqref{eq:29} vanishes. On the other hand, if point $\br$ is inside the causality sphere of one atom only, for example if $R_A < ct$ but $R_B > c t$, then the electric energy density in $\br$ is only related to the presence of atom $A$. This is compatible with relativistic causality, of course. Yet, interesting results are obtained if point $\br$ is inside the causality sphere of both atoms $A$ and $B$. In this case, all terms in \eqref{eq:29} contribute to the time-dependent electric energy density. In particular, assuming atoms with identical dipole moments, and for distances such that $R_A = R_B < ct$, we find that the electric energy density emitted by the two-atom system during the self-dressing process is doubled for atoms in the superradiant-state (with respect to the single-atom case), while it vanishes in the subradiant state. This is due to the presence of the interference term in \eqref{eq:29}, and clearly shows that the superradiant or subradiant behavior of the two-atom system can be understood, as far as the electric field energy density is concerned with, in terms of an interference effect between the electric energy densities emitted by the atoms.

We wish to point out that these features of the time-dependent electric energy density can be investigated through the retarded Casimir-Polder interaction energy with an appropriate polarizable body
with static polarizability $\alpha$. As it is known, in the far-zone limit, that is when the distance is larger than relevant wavelengths
associated to the atomic transitions, the Casimir-Polder energy can be written as
\begin{eqnarray}
\label{eq:31}
&\ & \Delta E = -\frac 1 2 \alpha \langle E^2(\br) \rangle ,
\end{eqnarray}
where the electric field operator ${\bf E} (\br)$ is evaluated at the position of the polarizable body.
Then, a third atom ($C$) located at some distance from atoms $A$ and $B$ feels a force that is directly related to the local electric field energy density generated at its position by the two entangled atoms. In particular, assuming that the distance of $C$ from $A$ and $B$ is such that $\mid \br_C-\br_A \mid=\mid\br_C-\br_B\mid$, the force on $C$ is enhanced if the two atoms are prepared in a correlated symmetric (superradiant) state, while it is zero if the atoms are prepared in a subradiant state, even if atom $C$ is inside the light-cone of both atoms $A$ and $B$.
Thus, the subradiant or superradiant behaviour of the two-atom system is strictly related to the field energy density, which can be experimentally probed through a measurement of the Casimir-Polder force on a third atom.

\section{\label{sec:4} Conclusions}
We have considered the time-dependent resonance interaction energy between two identical atoms, one in the ground state and the other in an excited state, interacting with the vacuum electromagnetic field, during the dynamical self-dressing process of the two atoms. In a quasi-static approach, it yields a resonance force between the atoms. This physical process is also strictly related to the resonant energy transfer between atoms. Using perturbation theory up to second-order, we have obtained the time-dependent  resonance interaction energy between the two atoms. We have shown that the resonance interaction vanishes when the two atoms are outside the light-cone of each other, while it settles to its stationary value immediately after the causality time $t=R/c$ (this is at variance of other dynamical Casimir-Polder interaction energies such as the dispersion atom-surface interaction \cite{VP08,MVP10,AVBBRP16}).  We have related these findings to a sort of interference between the contributions of real and virtual processes occurring during the time evolution of the system. We have also discussed  the time-dependent electric energy density in the space around the two entangled atoms, both for subradiant and superradiant states, and pointed out that its behavior is due to interference effects of the field emitted by the two atoms. Finally, we have shown that the behaviour of the time-dependent electric energy density for the two entangled atoms can be probed through the Casimir-Polder force on a third atom located in the vicinity of the two-atom system.

\acknowledgments
The authors gratefully acknowledge financial support from the Julian Schwinger Foundation. W. Zhou is thankful for financial support from the K. C. Wong Magna Fund in Ningbo University, and the research program of Ningbo University under Grant No. XYL18027.

\end{document}